\documentclass[a4paper,twocolumn,fleqn]{article}

\usepackage{amsmath}             
\usepackage{txfonts}               
\usepackage{cite}                  
\usepackage{graphicx}
\usepackage{pfr}                   

\begin{document}

\title{Formation of a Tungsten Co-deposition Layer with Microparticles Using Pulsed Laser Deposition}


\author{S.Kodate, Y.Hayashi, S.Kajita}

\affiliation{
  Graduate School of Frontier Sciences, The University of Tokyo, 5-1-5 Kashiwanoha, Kashiwa, Chiba 277-8561, Japan
}


\email{kajita@k-u.tokyo.ac.jp}

\begin{abstract}
  This study performed tungsten (W) co-deposition experiments using pulsed laser deposition under exposure to helium/argon plasma in a linear plasma device. 
  Micron-sized spherical W particles formed under co-deposition conditions. It was suggested that these 
  particles grew from nanoparticles via the electrostatic collection of W ions in the plasma plume before deposition.
\end{abstract}



\maketitle  



In magnetic confinement fusion reactors, plasma-facing materials (PFMs) such as divertor components must withstand extreme heat and particle fluxes.
Tungsten (W) is a leading PFM candidate due to its high melting point and low tritium retention.
It is planned for use in next-generation devices such as ITER~\cite{BOLT200243}.
However, W atoms can be released from the divertor surface via plasma-induced sputtering etc. and form a co-deposition layer~\cite{TOKITANI2024101678}. 
Co-deposition of W with other particles, such as hydrogen isotopes including tritium and helium (He) ash, can change the plasma material interaction and influence undesired tritium inventory~\cite{Roth_2008}, because it can significantly alter the material properties.
For example, it has been reported that a small W flux during He irradiation can accelerate the growth of nanostructures~\cite{Kajita2018EnhancedFuzz,Hori2023}, and the deposition layer drastically change the hydrogen retention behavior~\cite{ALIMOV2010225,KAJITA2020152350}.
However, the formation and behavior of co-deposition layers with transient plasma events such as edge-localized modes (ELMs)~\cite{LEONARD1999109, Loarte_2003}, which can impose a transient heat load of several GW/m$^2$ within $0.1-1 \ \mathrm{ms}$, are almost unknown.
It is interesting to investigate whether the formed co-deposition layer changes under transients, because the deposition rate of W can be higher over a shorter time period than under steady-state deposition conditions.


This study combined a linear plasma device with pulsed laser deposition (PLD) to form co-deposition layer with transient W deposition under plasma exposure.
Microparticles are shown to form on the substrate. The size distribution and formation mechanism of the microparticles are discussed based on surface observations.

\begin{figure}[tb]
  \centering
  \includegraphics[width=7cm]{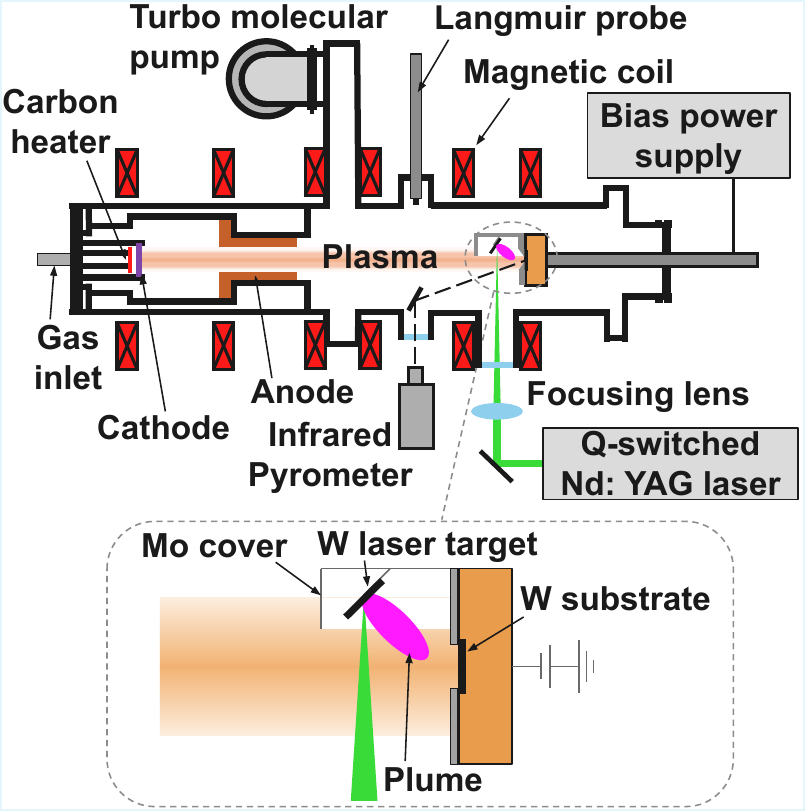}
  \caption{Schematic of the experimental setup}
  \label{fig1}
\end{figure}

\begin{figure*}[t]
  \centering
  \includegraphics[width=0.8\textwidth]{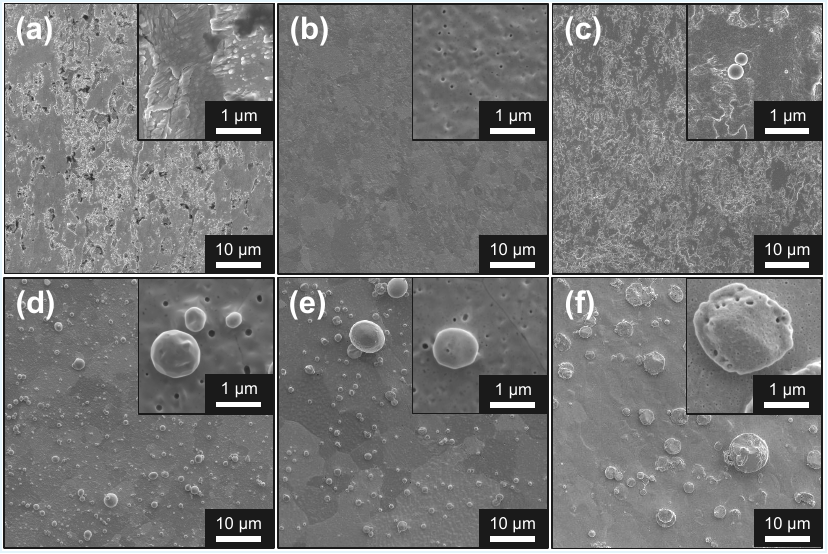}
  \caption{SEM micrographs of the W substrate surface: (a) pristine substrate; (b) plasma irradiation only (incident ion energy: 16 eV); (c) W deposition only; and (d-f) co-deposition at incident ion energies of 16 eV, 21 eV, and 31 eV, respectively.}
  \label{fig2}
\end{figure*}

Experiments were conducted using a linear plasma device (LOTUS-II) in combination with a Q-switched Nd:YAG laser 
(Continuum Surelite II), as shown in Fig.~\ref{fig1}.
LOTUS-II was developed by relocating and upgrading NAGDIS-I from Nagoya University~\cite{Masuzaki_1990}.
Plasma was generated via direct current (DC) arc discharge in a He/argon(Ar) gas mixture (He: $4.8 \ \mathrm{Pa}$; Ar: $0.3 \ \mathrm{Pa}$) within a magnetic field of $\sim0.03 \ \mathrm{T}$ 
and was directed toward a W substrate ($20 \times 20 \times 0.2 \ \mathrm{mm}$). The pristine substrate was cleaned with ethanol and left unpolished.
A small amount of Ar was added to enhance the ionization efficiency.
In the divertor region, Ar injection is also used to lower the plasma temperature and heat flux by promoting radiation cooling and volume recombination~\cite{FOURNIER1998231}. Therefore, the potential impact of the introduced Ar on the substrate structure should be considered.
Simultaneously, the pulsed laser (wavelength: $532 \ \mathrm{nm}$; pulse width: $4-7 \ \mathrm{ns}$; repetition rate: $10 \ \mathrm{Hz}$) was focused on a W laser target 
(diameter: $30 \ \mathrm{mm}$; thickness: $1 \mathrm{mm}$; purity: $99.9\%$), which is located near the W substrate, to ablate material and form a plume. 
This plume was then deposited on the substrate to fabricate a co-deposited layer.
The $7 \ \mathrm{mm}$ diameter laser beam was focused to a spot area of approximately $3.9 \ \mathrm{mm^{2}}$ on the laser target, corresponding to a laser fluence of  $\sim 5.0 \ \mathrm{J/cm^{2}}$.
The W laser target was positioned about $4 \ \mathrm{cm}$ from the substrate and tilted at $45^\circ$.
The W substrate was mounted on a water-cooled copper plate and biased to control the incident ion energy.
The plasma parameters measured at the plasma center using a Langmuir probe were as follows:
electron density $n_e \approx 2.8 \times 10^{19} \ \mathrm{m^{-3}}$, electron temperature $T_e \approx 4.0 \ \mathrm{eV}$, and ion 
flux $\Gamma_i \approx 1.2 \times 10^{23} \ \mathrm{m^{-2}s^{-1}}$.
The substrate surface temperature was monitored using an infrared pyrometer (Lec, KTL-PRO, wavelength: 1.6 \textmu m, emissivity: 0.3) through an observation window, and remained steady at $1080-1220 \ \mathrm{K}$ under both co-deposition and irradiation with plasma only conditions.

The mass of each sample was measured before and after the experiments using a microbalance (A\&D, BM-20). The difference in the areal mass change was then obtained.
Surface morphology was observed using a scanning electron microscope (SEM, JEOL JSM-7001F).
To quantify the size distribution of deposited particles, SEM micrographs were processed using the following steps: noise reduction, edge detection, and binarization to extract particle regions, and watershed segmentation to separate individual particles.
Labeled-region analysis of the micrographs was then used to calculate the particle size distribution based on particle diameters.

Under the present co-deposition conditions, the plasma directly heats the laser target, which may have influenced to the experimental results.
Therefore, as shown in Fig.~\ref{fig1}, a molybdenum (Mo) cover was placed on the laser target to investigate microparticle morphology as a function of the laser target temperature ($T_L$).
This temperature was also measured through the observation window using the infrared pyrometer.

\begin{figure*}[t]
  \centering
  \includegraphics[width=1.0\textwidth]{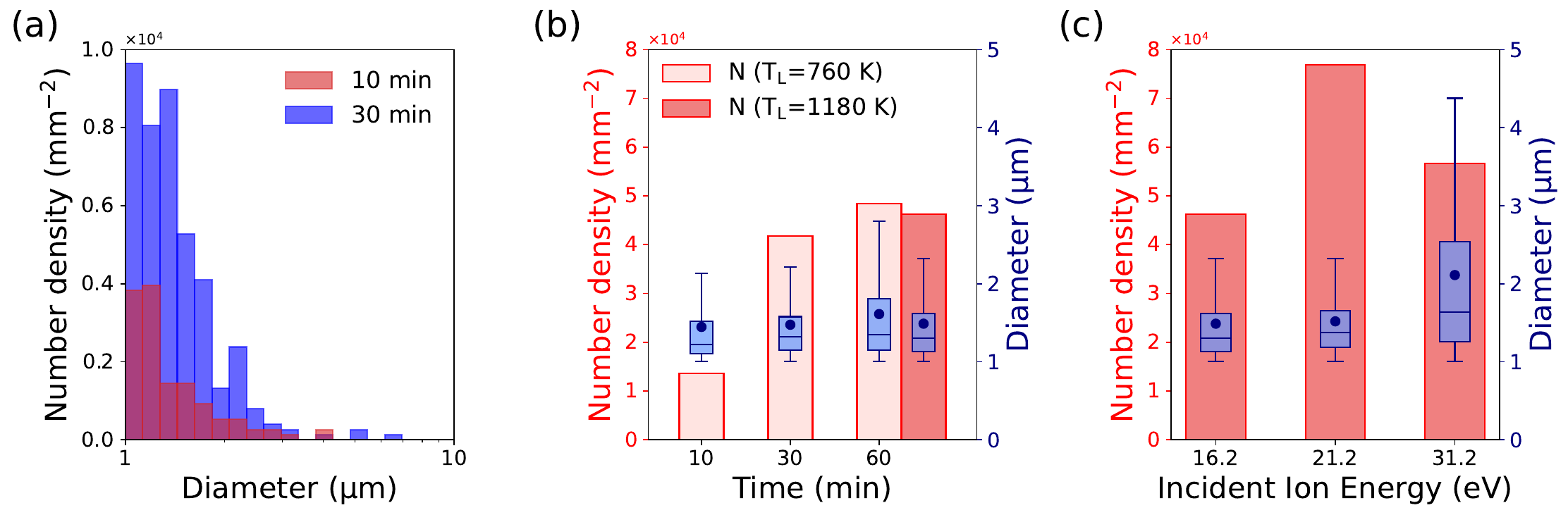}
  \caption{(a)Size distributions of microparticles for deposition times of 10 and 30 minutes. (b)Total number of microparticles and their size distributions for different deposition times and $T_L$. (c)Total number of microparticles and their size distributions at different incident ion energies.}
  \label{fig3}
\end{figure*}

Fig.~\ref{fig2}(a) shows an SEM image of a pristine W substrate with a relatively uneven structure, but no micro-particles were identified.
Fig.~\ref{fig2}(b) shows the morphology of the substrate surface after 60 min of plasma irradiation at the incident ion energy of $16 \ \mathrm{eV}$.
The surface appeared flattened, with dark dot-like structures and voids that were several tens of nanometers in size.
These pinholes are attributed to the growth of He bubbles due to plasma irradiation \cite{NISHIJIMA20041029, 10.1063/1.2824873, CIPITI2005298}.
The mass of the substrate decreased by $0.07 \ \mathrm{mg/cm^2}$ due to the sputtering by a small amount of Ar ions, suggesting that 30-40~nm of erosion occurred during irradiation.
Figure~\ref{fig2}(c) shows an SEM micrograph of a W surface with 60~min W deposition without plasma irradiation. 
The morphology of the substrate surface with the W deposition was relatively uniform, and nanoparticles ranging in diameter from $10 \ \text{to} \ 100 \ \mathrm{nm}$ were observed.
In Fig.~\ref{fig2}(d), after 60 min of co-deposition with the laser and plasma at the incident ion energy of $16 \ \mathrm{eV}$, notably larger micron-sized spherical W particles (microparticles) 1 \textmu m in diameter were observed across the surface.
In addition, pinholes can be identified on the surface, similar to the condition with plasma-only irradiation, suggesting that the He bubbles were formed.
The mass of the substrate increased by 0.14 and $0.16 \ \mathrm{mg/cm^2}$, in Fig.~\ref{fig2}(c) and (d), respectively.
Fig.~\ref{fig2}(e,f) show SEM micrographs of the co-deposition layer with higher incident ion energies of 21 and 31~eV, respectively. 
At $21 \ \mathrm{eV}$, the microparticles exceeding 1 \textmu m in diameter were deposited similarly, while the number of particles smaller than 1 \textmu m in diameter decreased.
The pinholes were generally smaller in size than those at $16 \ \mathrm{eV}$.
At $31 \ \mathrm{eV}$, microparticles were still deposited, but appeared compressed and less rounded.
The pinholes were scarcely observed on the flat surface, possibly due to the sputtering of the He bubble layer by Ar ions.
It is noted that the He bubbles were observed on the deposited microparticles.
In Fig.~\ref{fig2}(e,f), the mass of the substrate decreased by 0.07 and $0.77 \ \mathrm{mg/cm^2}$, respectively, suggesting that deposition and sputtering occur simultaneously.
In particular, at 31~eV, which is slightly greater than the sputtering threshold energy~\cite{Eckstein2002IPP},
the mass change indicated that $\sim$400~nm of W was sputtered during the exposure. 
The induced sputtering process likely altered the shape of the microparticles at 31~eV. 
In contrast, at 16 eV---below the sputtering threshold of 28 eV for Ar on W---mass measurements confirmed that deposition dominated over sputtering.

In Fig.~\ref{fig3}(a), the size distributions of the microparticles for deposition times of 10 and 30~min at $16 \ \mathrm{eV}$ are shown. It is noted that the minimum size threshold was set at the diameter of 1~$\mu$m, because noise can be detected as particles when the threshold is decreased further. 
For particles between 0.2 and 1~$\mu$m in diameter, the number density at $T_L=760$~K was estimated to be 0.5$\times10^5$~mm$^2$ after 10 min. The number density of the small particles ($< 1~\mu$m) is greater than that of the larger particles ($>1~\mu$m) and increases with time. However, as the deposition time increased, the growing surface roughness of the substrate made it increasingly difficult to distinguish particles smaller than 1~$\mu$m. Therefore, in the present study, we focused on particles larger than 1~$\mu$m, for which the influence of noise was negligible.
In Fig.~\ref{fig3}(a), the microparticles were primarily 1-3~$\mu$m in size.
Although the number of microparticles increased with the deposition time, the distribution did not change significantly.

Fig.~\ref{fig3}(b) summarizes the total number of microparticles and size distributions of different deposition times at $16 \ \mathrm{eV}$.
As seen in Fig.~\ref{fig3}(a), the number of particles increased over time. However, the size distribution showed no significant change, and the average diameter remained nearly constant.
The lack of change in the size distribution of the microparticles over deposition time indicates that they grew before reaching the substrate, rather than grown on the substrate.

In Fig.~\ref{fig3}(b), at 60~min, by removing the Mo cover, a comparison was made between $T_L$=760 and 1180~K. 
Between the two conditions, only minor changes were observed in the number of microparticles and in their size distribution in Fig.~\ref{fig3}(b). 
The mass increase was $0.17 \ \mathrm{mg/cm^2}$ at $760 \ \mathrm{K}$ and $0.16 \ \mathrm{mg/cm^2}$ at $1180 \ \mathrm{K}$, showing nearly the same values.
This comparison suggests that the observed microparticles were not formed by direct droplet ejection caused by laser ablation.
Instead, they were likely formed through interactions between W particles and plasma species during transport in the plume.

Fig.~\ref{fig3}(c) shows the total number of microparticles and their size distribution for different incident ion energies with 60 min deposition time.
The number of microparticles increased from $16 \ \mathrm{eV}$ to $21 \ \mathrm{eV}$ but decreased at $31 \ \mathrm{eV}$, likely due to Ar-induced sputtering of deposited microparticles.
The size distribution between $16 \ \mathrm{eV}$ and $21 \ \mathrm{eV}$ showed no significant differences, whereas at $31 \ \mathrm{eV}$ the distribution shifted toward larger diameters, with the average size exceeding 2 \textmu m only under this condition.



Under low-pressure PLD conditions (a few Pa), cluster growth via gas-phase condensation is generally suppressed \cite{KEK2020137953, Kodate_2024}.
Moreover, negatively charged particles in plasma experience mutual Coulomb repulsion, which inhibits aggregation.
However, in this study, such microparticles formed only under co-deposition conditions, suggesting a growth mechanism that involves interactions between negatively charged W nanoparticles and W ions in the plasma plume.
We believe that sputtering erosion is unlikely to play a significant role under in the present experiments, because no change in size distribution change was observed at an incident ion energy of 16~eV, which is much lower than the sputtering threshold.
One plausible mechanism is that negatively charged W nanoparticles interact with W ions in the plasma plume.
The plume produced by the laser pulse is known to contain W atoms, W ions, electrons, and W nanoparticles (tens of nm)~\cite{KAJITA2025102529}.
Due to their large surface-to-volume ratio, nanoparticles can efficiently capture electrons and become negatively charged, thereby potentially attracting positively charged W ions. Repeated ion collection could lead to particle growth.
In the early stage of plume formation, the electron density is estimated to be $n_e \approx 10^{23} \, \mathrm{m}^{-3}$, with electron temperature $T_e \approx 1 \, \mathrm{eV}$ and ion temperature $T_i = 0.1 \sim 1 \, \mathrm{eV}$ \cite{ESCALONA2021104066}.
Electrons, being much faster than ions, charge the nanoparticles within nanoseconds.
Within several hundred nanoseconds to microseconds, the plume relaxes into the background plasma with $n_e \approx 2.8 \times 10^{19} \, \mathrm{m}^{-3}$ and $T_e \approx 4.0 \, \mathrm{eV}$.
Consequently, the Debye length increases from approximately $24 \ \mathrm{nm}$ to 2.8 \textmu m, which may enhance W-ion collection by nanoparticles.
In addition, the relatively long persistence of W ions in the plume could sustain their interaction with charged nanoparticles. In contrast, during W deposition without plasma, the lower $n_e$ suppresses nanoparticle charging, which may in turn limit ion collection and the subsequent growth of microparticles. Although further experimental validation is required, this mechanism provides a consistent explanation for the observed microparticle formation under co-deposition conditions.

Here, to estimate the growth rate of W nanoparticle diameter, the Orbital Motion Limited (OML) theory \cite{PhysRev.28.727} was applied.
It was assumed that W ions and plume electrons flowing into the W nanoparticle follow a shifted Maxwellian distribution in a coordinate system centered on the W nanoparticle.
Given the directional nature of the ablation plume toward the substrate, W ions and plume electrons are considered to flow predominantly into the nanoparticle from one direction.
Assuming that W ions are singly charged, the potential of the W nanoparticle relative to the plasma potential is estimated to be approximately $-4 \ \mathrm{V}$, determined from the condition that the net incoming current of W ions and electrons is zero.
Based on this charging potential, the flux of W ions to the nanoparticle was calculated.
Assuming a transport time of several hundred microseconds from the laser target to the substrate, a nanoparticle with an initial diameter of $10 \ \mathrm{nm}$ is expected to grow to several hundred nanometers to one micrometer by the time it reaches the substrate.
This estimate is consistent with the trend observed in the experimental results.

In summary, the W co-deposition experiments were conducted using PLD under He/Ar plasma irradiation to simulate ELM-like W co-deposition in LOTUS-II.
Micron-sized spherical W particles were formed only under co-deposition conditions.
SEM analysis and growth rate estimation suggest that these microparticles grew in the plasma plume via electrostatic collection of W ions before deposition rather than surface growth or droplet ejection.
These results imply that in fusion reactor divertors, W nanoparticles released during transient high-flux events such as ELMs can grow via ion collection in the plasma and eventually redeposit as micron-sized dust particles on the divertor surface. \\

The authors thank Dr.~K. Murakami of the University of Tokyo for permission to use the SEM.
This work was supported in part by the Grant-in-Aid for Scientific Research by the Japan Society for the Promotion of Science (JSPS) under Grant Nos. 21KK0048, and 25H00615. This work was performed with the support and under the auspices of the NIFS Collaboration Research Program (NIFS25KFGT001 and NIFS25KRCT001). 


\end{document}